\documentstyle[12pt]{article}
\input epsf.tex
\newcommand{\veck}{{\bf k}}
\newcommand{\vecs}{{\bf s}}
\newcommand{\vecL}{{\bf L}}
\newcommand{\vecA}{{\bf A}}
\newcommand{\vecp}{{\bf p}}
\newcommand{\vecv}{{\bf v}}
\newcommand{\vecr}{{\bf r}}
\newcommand{\vecQ}{{\bf Q}}

\newcommand{\vecE}{{\bf E}}
\newcommand{\vecB}{{\bf B}}
\newcommand{\vecD}{{\bf D}}

\newcommand{\vecsigma}{\mbox{\boldmath $\sigma$}}
\def \pdag{\psi'^{\dagger}}

\def \sl#1 {\not \! #1}

\newcommand{\beq}{\begin{equation}}
\newcommand{\eeq}{\end{equation}}
\newcommand{\beqa}{\begin{eqnarray}}
\newcommand{\eeqa}{\end{eqnarray}}
\newcommand{\beqar}{\begin{eqnarray*}}
\newcommand{\eeqar}{\end{eqnarray*}}
\renewcommand{\a}{\alpha}
\renewcommand{\d}{\delta}
\newcommand{\D}{\Delta}

\begin{document}
\begin{titlepage}
\rightline{\small  \hfill McGill/96-41}
\vskip 5em

\begin{center}
{\bf \huge 
 
Derivation of the Lamb Shift \\[.25em]using an Effective Field Theory}
\vskip 3em

{\large Patrick Labelle \footnote{labelle@hep.physics.mcgill.ca}}
\vskip 1em
{\em \it{ Department of Physics, Bishop's University \\ Lennoxville,
Qu\'ebec, Canada J1M 1Z7}}
\vskip 1em
and \\
\vskip 1em
{\large S. Mohammad Zebarjad\footnote{zebarjad@hep.physics.mcgill.ca}}
\vskip 1em

{\em \it{	Department of Physics, McGill University \\
Montr\'eal, Qu\'ebec, Canada H3A 2T8}}
\vskip 4em

\begin{abstract}
We rederive the ${\cal O} ( \alpha^5)$ shift of the hydrogen
levels in the non-recoil limit  ($m_e/m_P \rightarrow 0$)
  using Nonrelativistic QED (NRQED),
an effective field theory developed by Caswell and Lepage
 (Phys. Lett.{\bf 167B}, 437 (1986)). Our result contains the
 Lamb shift as a special case. Our calculation is 
far simpler than traditional approaches  and has the
advantage of being systematic.
It also clearly illustrates 
 the need to renormalize (or ``match")
 the coefficients of the
effective theory beyond tree level.
\end{abstract}
\end{center}

\end{titlepage}

\setcounter{footnote}{0}
\section{Introduction}

The Lamb shift, 
 the shift between the hydrogen  $ 2S_{1/2}$ 
and $2P_{1/2}$ states, is without any  doubt the most well-known bound state
application of radiative corrections.
It has the form $ m_e \alpha^5 \big( \ln \alpha +{\rm
finite} \bigr)$ where the finite piece contains the Bethe logarithm, 
a state dependent term that must be evaluated numerically. The log term
is relatively easy to extract and its calculation is presented in many
quantum mechanics textbooks. The finite contribution is much more
difficult to evaluate because it requires the application of QED to a
bound state. In this paper, we rederive the complete ${\cal O} ( 
\alpha^5)$ corrections in the non-recoil limit\footnote{In the
literature, these
corrections are also
 referred to as the Lamb  shift. We will also adopt this
notation in the rest of the paper.} ($m_e/m_P =0$) using
NRQED,  an effective
field theory  developed by Caswell and Lepage \cite{Lepage}, as extended
 by Labelle \cite{MQED} to study retardation effects.

To construct NRQED, one writes down the most general Lagrangian
consistent with the low energy symmetries of QED such as parity and
gauge invariance, etc. The first few terms of this Lagrangian are given
by (we follow the notation of \cite{kinoshita})
\beqa
{\cal L}_{2-Fermi}~& =  &\psi^{\dagger}\{ iD_{t} +
\frac{\vecD^2}{2m}
+\delta_R \frac{\vecD^4}{8m^{3}}
+ \delta_{F}\frac{q \vecsigma\cdot\vecB}{2m}
+ \delta_{D}\frac{q (\vecD\cdot\vecE -\vecE\cdot\vecD)}{8m^{2}}
 \nonumber \\
& & +\delta_{S}\frac{iq \vecsigma\cdot(\vecD\times \vecE
  - \vecE\times \vecD )}{8m^{2}}+ \dots \}\psi_e \nonumber \\
& & ~+~ {\rm same~terms~with~} \psi_e \rightarrow \psi_p  
~+~ {\rm photon ~terms}, \label{lag}
\eeqa
where $\psi_e$ and $\psi_p$ are two component fields associated with the
electron and proton, respectively.
 There are of course many
other interactions, including four-fermion interactions, however, as we
will discuss below, those are not relevant at
 ${\cal O} ( \alpha^5) $.
 For the photon, we use the Coulomb gauge which is the most efficient
gauge to study nonrelativistic bound states since it permits to isolate
the Coulomb interaction (which must be treated nonperturbatively) from
all other interactions (which can be treated as perturbations). In that
gauge, the photon terms in Eq.(\ref{lag})  are 
\beqa
-{1 \over 4 } F_{\mu \nu} F^{\mu \nu} + \delta_{VP}  { \alpha \over 15 \pi}
 A^0 (\veck) { \veck^4
\over m^2} A^0 (\veck)  - \d_{VP} { \alpha \over 15
\pi}  A^i(k) {\veck^4 \over m^2} A^j (k)
~(\delta_{ij} - { k_i k_j \over \veck^2} ) + \dots \label{photonlag}
\eeqa

\begin{figure}
\centerline{\epsfxsize 5.2 truein
\epsfbox {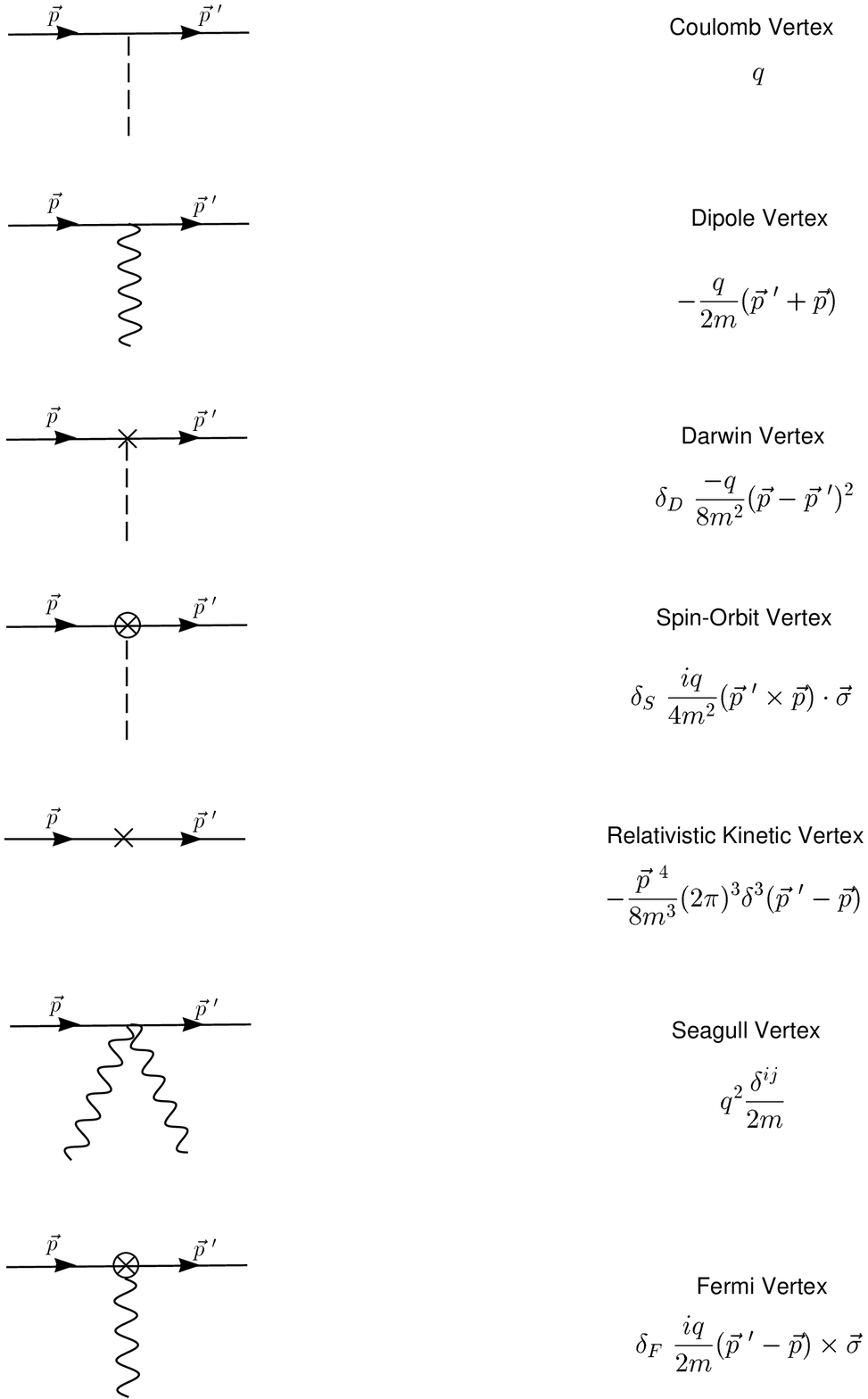}}
\caption{NRQED Vertices }
\label{fvertices}
\end{figure}

Using $ \vecD = i(\vecp- q \vecA) $ and $D_t= \partial_t+iqA_0$, the
 NRQED hamiltonian can be written as
\beqa
{\cal H}=\pdag\biggl[ {\vecp^2\over 2m}&+&qA_0-{\vecp^4\over 8m^3}-
{q\over 2m}(\vecp'+ \vecp) \cdot \vecA + {q^2\over2m}
\vecA \cdot \vecA \nonumber\\
&& -\d_F{iq\over2m}
\vecsigma \cdot( \veck \times \vecA)-\d_D{q\over 8m^2}\veck^2 A^0+
\d_S{iq\over 4m^2} \vecsigma \cdot(\vecp' \times \vecp)A^0\nonumber\\
&&+\d_S{iq\over8m^2}k^0\vecsigma \cdot (\vecp'+\vecp)\times \vecA
-\d_S{iq\over4m^2}\vecsigma\cdot(\veck_1 \times \vecA(k_1))A^0(k_2)+
\dots\biggl]\psi,  \nonumber\\ \label{hamiltonian}
\eeqa
and the photon hamiltonian can be written as
\beqa
{\cal H}_{photon}={1\over 2}(\vecE^2+\vecB^2)-\d_{VP}{\a \over 15 \pi}
 A^0(\veck){\veck^4\over m^2} A^0(\veck)+\d_{VP}{\a \over 15 \pi}A^i(k){\veck^4\over m^2} A^i(k) 
(\delta_{ij} - { k_i k_j \over \veck^2} )+\dots\nonumber\\
\label{hamphoton}
\eeqa
The Feynman
 rules of Eqs.(\ref{hamiltonian}) and  (\ref{hamphoton}) are given in
 Fig.[\ref{fvertices}]. 

\begin{figure}
\centerline{\epsfxsize 5.2 truein
\epsfbox {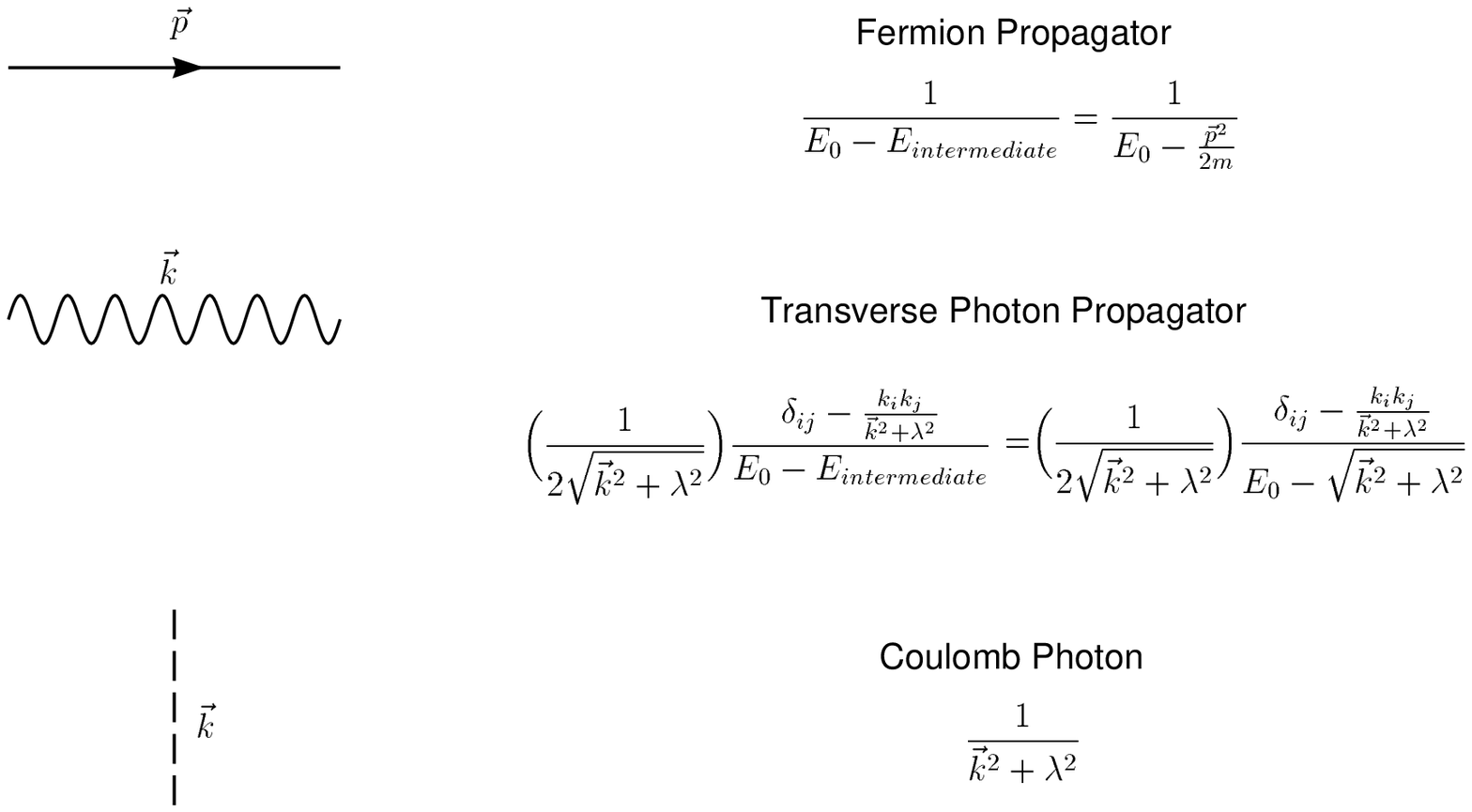}}
\caption{NRQED Propagator }
\label{fpropagator}.
\end{figure}

\begin{figure}
\centerline{\epsfxsize 5.2 truein
\epsfbox {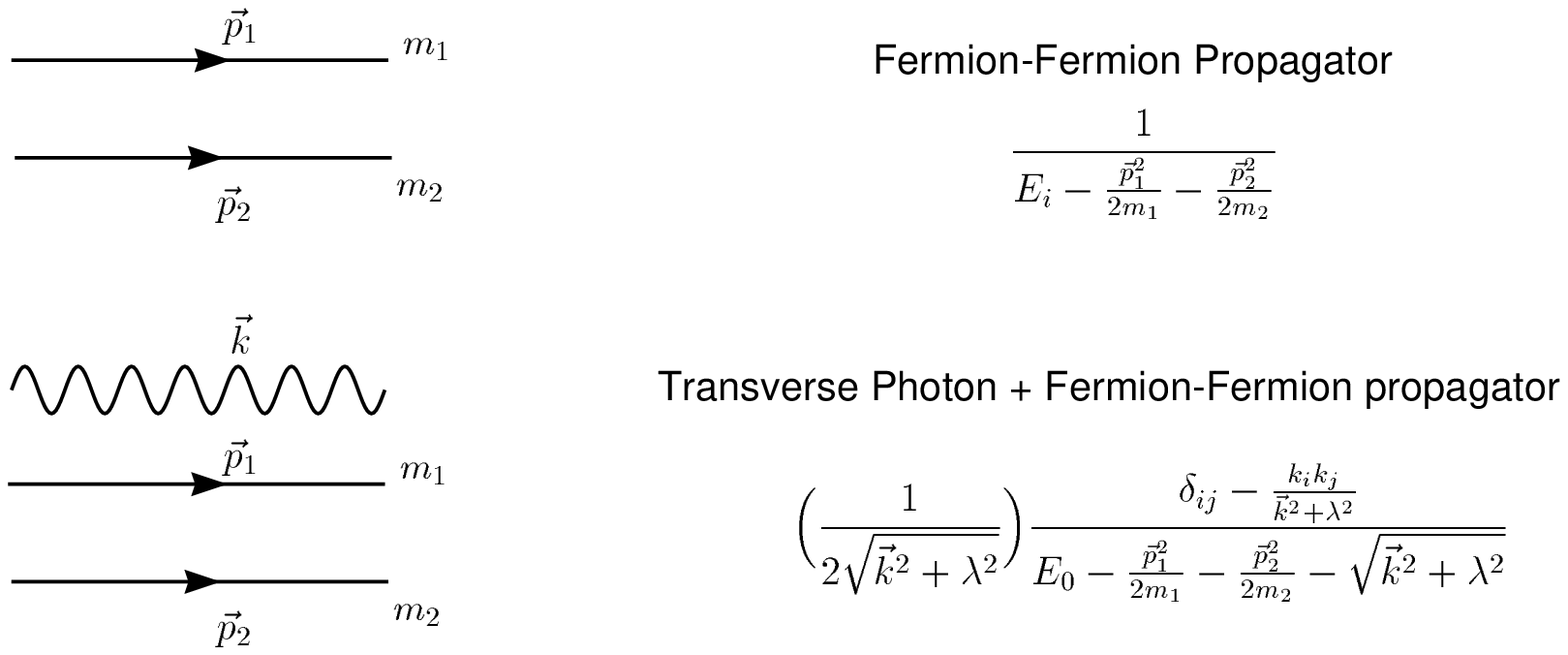}}
\caption{Time-ordered propagators for two fermions plus one
transverse photon. }
\label{fpropagators}.
\end{figure}

The photon propagator   requires a special treatment.
 We use time-ordered perturbation  theory (in which all
particles are on-shell but energy is not conserved at the vertices)
and, as mentioned above, we work in the Coulomb gauge.
In  a time-ordered diagram, Coulomb photons  propagate
instantaneously  ({\it i.e.} along vertical lines in our diagrams
since we choose the time axis to point to the right)
and   transverse photon 
 propagate in the time direction.  The corresponding Feynman  rules  are given
in Figs.[\ref{fpropagator}] and 
[\ref{fpropagators}].

\begin{figure}
\centerline{\epsfxsize 5.2 truein
\epsfbox {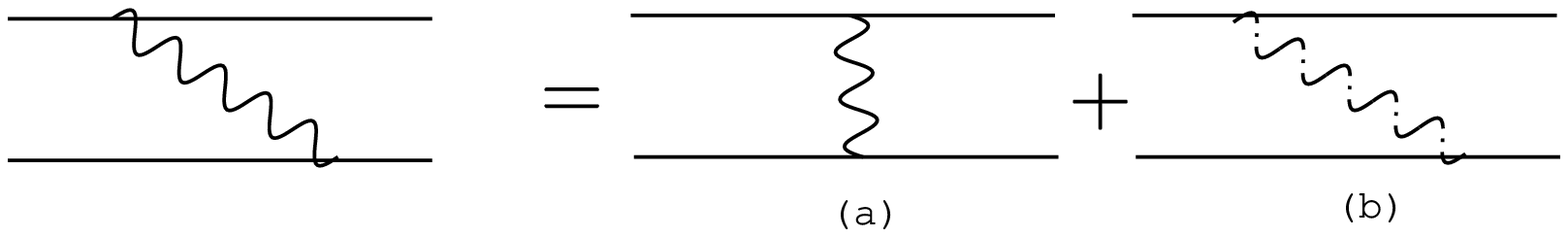}}
\caption{ }
\label{fsoft-ultrasoft}.
\end{figure}

As is shown in \cite{MQED}, the power of NRQED is enhanced if one
 separates ``soft" transverse photons (having energies
of order $ \gamma \equiv Z \mu \alpha$ ) from the ``ultra-oft" photons
(with $E \simeq \gamma^2/ \mu$) because the counting rules differ for
the two types of photons.
 In that reference, it is also shown that
the multipole expansion can be applied to the vertices containing
ultra-soft photons.  The distinction between soft and ultra-soft photons
is particularly important in this work because the Lamb shift involves
both types of  contributions.  We have therefore separated the general
photon propagator of Fig.[\ref{fsoft-ultrasoft}] into a contribution
 corresponding
 to a soft photon, Fig.[\ref{fsoft-ultrasoft}(a)], and the contribution 
corresponding to an  ultra-soft photon, Fig.[(\ref{fsoft-ultrasoft}b)].

Finally, in NRQED loop diagrams, all internal 
 momenta must be integrated over, with a measure  $d^3p /(2 \pi)^3$.

The  unknown coefficients $\d_R, \d_F \ldots$  can be found by imposing that
NRQED is equivalent to QED for  nonrelativistic scattering 
processes, 
 {\it i.e.} by imposing \\
\begin{center} 
\begin{minipage}{2in}
\begin{center}  NRQED scattering amplitudes\\ 
   \end{center}
\end{minipage}\ {\Large = } \
  \begin{minipage}{2in} \begin{center}QED scattering amplitudes
\\  expanded 
 in powers of $ \vecp/m$.  \end{center}
\end{minipage}  
 \end{center}
This is the so-called matching procedure. Notice that no
 bound state physics enters at this stage of the calculation.

\begin{figure}
\centerline{\epsfxsize 5.2 truein
\epsfbox {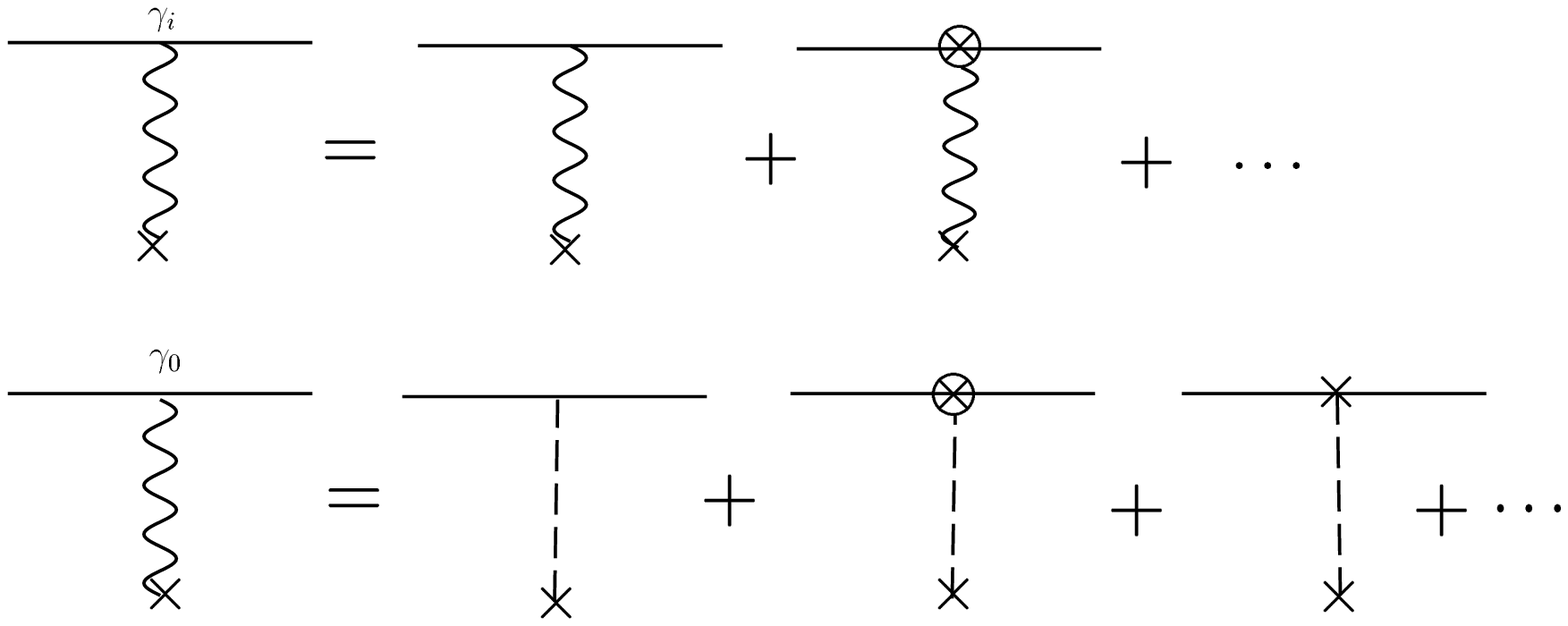}}
\caption{ }
\label{ftree-matching}.
\end{figure}

\begin{figure}
\centerline{\epsfxsize 5.2 truein
\epsfbox {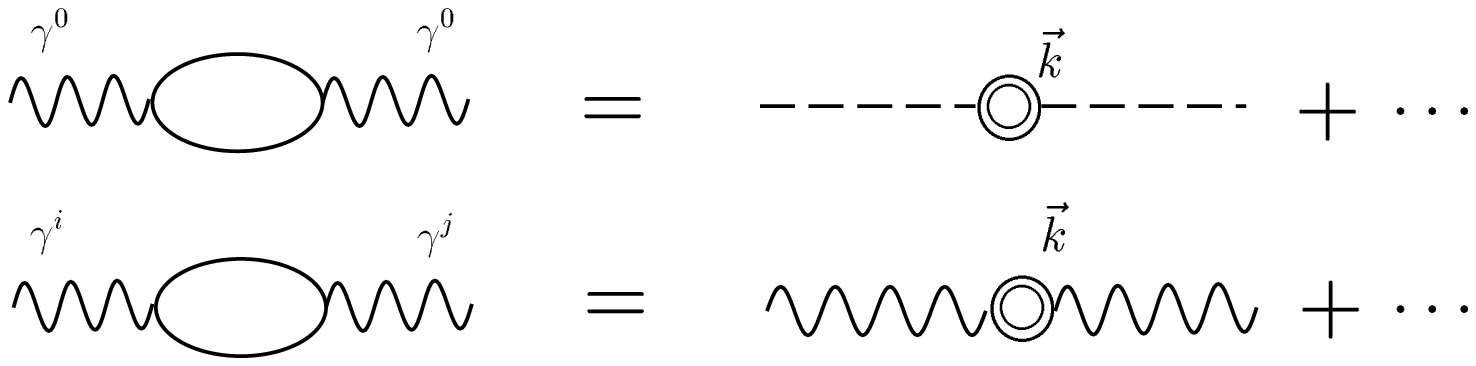}}
\caption{ }
\label{fvacuum-matching}.
\end{figure}

The coefficients appearing in (\ref{hamiltonian}) can be fixed
by considering the scattering of an electron off 
 an external 
  field $A_{\mu}$. This is illustrated, at tree level,
 in Fig.[\ref{ftree-matching}].
We have chosen the normalization of the interactions of
Eq.(\ref{hamiltonian}) in such a way that 
 the tree level matching gives $\delta=1$ for
all the coefficients appearing in (\ref{hamiltonian}). The first
non-zero contribution to $\delta_{VP}$ comes from the one-loop
QED vacuum polarization diagram, as illustrated in
 Fig.[\ref{fvacuum-matching} ]
and, again, our normalization is such that $\d_{VP} =1$,
to one loop \cite{kinoshita}. We will
still refer to this as ``tree level matching" because only tree
level NRQED diagrams are involved
(similarly, n-loops matching will refer to the number of loops ``n"
 the NRQED diagrams).
 The one-loop matching will bring ${\cal O} (\alpha)$
corrections to the coefficients $\delta's$.

We are now in position to proceed with the calculation.
 All NRQED calculations can be divided into three
steps. The first one consists in using the counting rules
to identify the diagrams which will contribute to the order
of interest. This first step not only permits to identify the
relevant diagrams but it also fixes the order (in the number of
loops) at which the coefficients must be matched. The second step
consists in matching the coefficients to the order required
and the last step corresponds to finally evaluating the NRQED
bound state diagrams.

\begin{figure}
\centerline{\epsfxsize 5.2 truein
\epsfbox {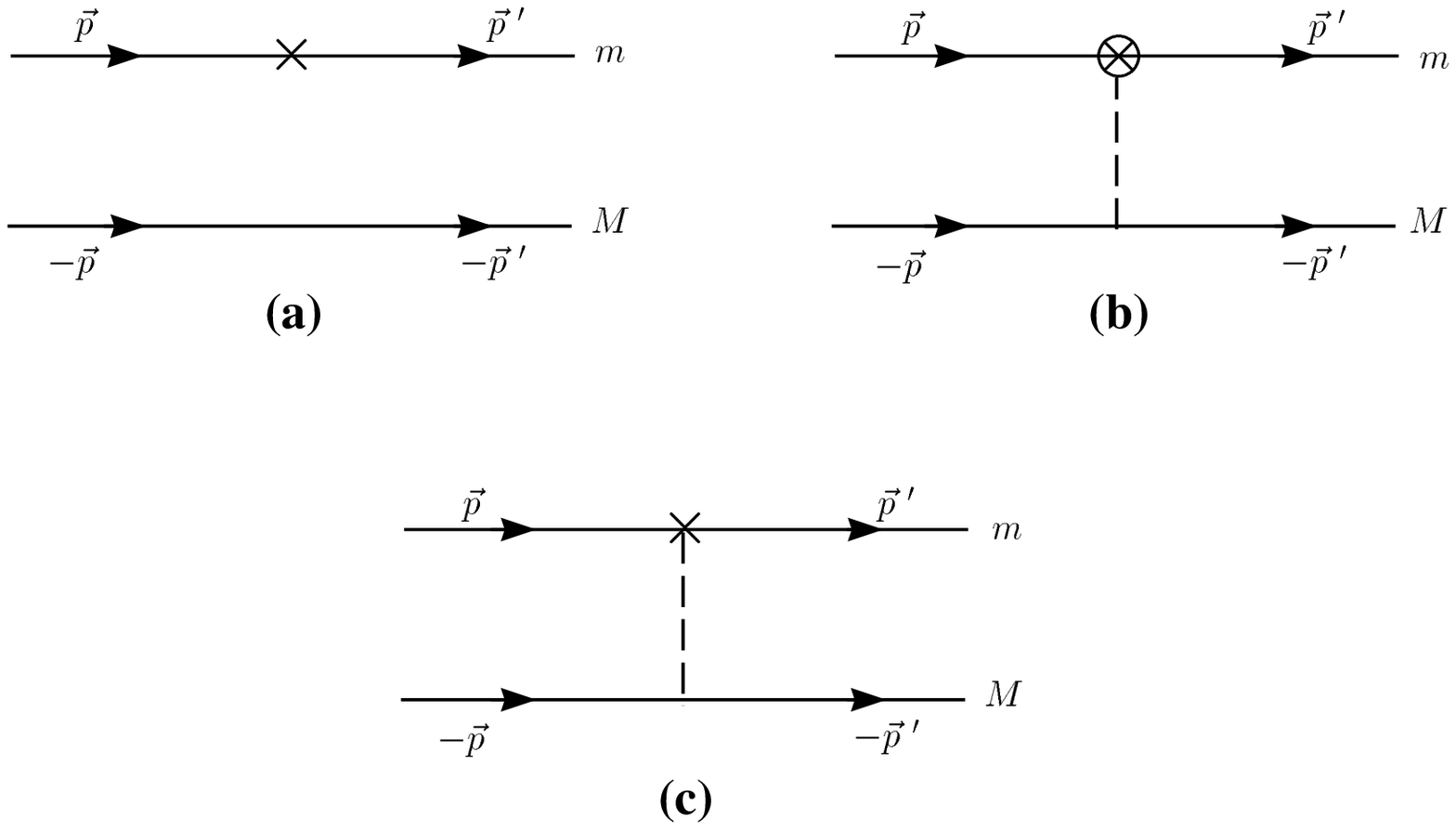}}
\caption{ }
\label{a4}.
\end{figure}

In our case,  we first need to isolate the NRQED diagrams
contributing to order $  \alpha^5$ in the non-recoil limit
{\it i.e.} we neglect corrections suppressed
by powers of $m_e / m_p$.
   As a simpler example, we will first isolate the
diagrams contributing to order $ \alpha^4$ in the non-recoil limit
 (the full NRQED calculation of the ${\cal O} (\alpha^4) $  
energy shift
 for arbitrary masses 
will be  presented in Ref.\cite{pat-moh}). In that limit, the only 
relevant diagrams
  are shown in Fig.[\ref{a4}]. 
 This can  easily be checked by
using the NRQED counting rules derived in Ref.[2]. Since 
soft and ultra-soft photons obey different counting rules, we 
consider their contribution in turn.   Soft photons  
contribute 
to order :
\beqa
{\mu^{\kappa + \rho + 1} \over {m_e^\kappa m_p^\rho} } Z^\eta \alpha^\zeta
\approx  {m_e^{\rho+1} \over m_p^\rho}  \alpha^\zeta, \label{crules}
\eeqa
where $\mu$ is the reduced mass and  
 $\rho$ and $\kappa$ are, respectively,
 the total number of  inverse powers of   $ m_p$ and $ m_e$ 
appearing in the NRQED vertices. In (\ref{crules}), we approximated
 $\mu \approx m_e$ which is exact in the non-recoil limit.
 The coefficient $ \zeta$ is
 defined by
\beqa
\zeta=1+\kappa +\rho -N_{TOP}+\sum_i n_i,
\label{zeta}
\eeqa
where $N_{TOP} $ is the number of intermediate state time-ordered
 propagators (see Ref.[2] for more details) and
last term is the sum of factors of $\a$ contained in the vertices
 (including
 possible factors coming from  the $\d$'s). 
Now, to obtain the correction of order $m_e \alpha^4$, we need to have
$\rho =0$ and $\zeta=4$ (see Eq.(\ref{crules})).
  The only way to have  $\rho =0$ is to either have
 a Coulomb interaction on the nucleus line or no interaction at
all.  This already reduces greatly the possible diagrams.
Turning now to the condition  $\zeta =4$, we obtain from Eq.(\ref{zeta})
\beqa
\kappa- N_{TOP} + \sum_i n_i = 3 .
\eeqa

 Since in first order of perturbation
 theory $N_{top}=0$,
we are left with the condition 
  $\kappa+\sum_i n_i=3$. One possibility is $\kappa =3 $ and
$ \sum n_i =0$ which corresponds to the relativistic kinetic energy
vertex on the electron line. Another possibility is $\kappa=2$ and $\sum
a_i = 1$ which can be fulfilled with  the Coulomb vertex on the proton
line and either the Darwin or the Spin-Orbit interaction on the 
electron
line. There are no  Coulomb interaction with only one inverse mass so
the condition $\kappa =1 $ and $ \sum n_i =2$  cannot be satisfied. The 
three possible diagrams are illustrated in Fig.[\ref{a4}].
 
\begin{figure}
\centerline{\epsfxsize 2.8 truein
\epsfbox {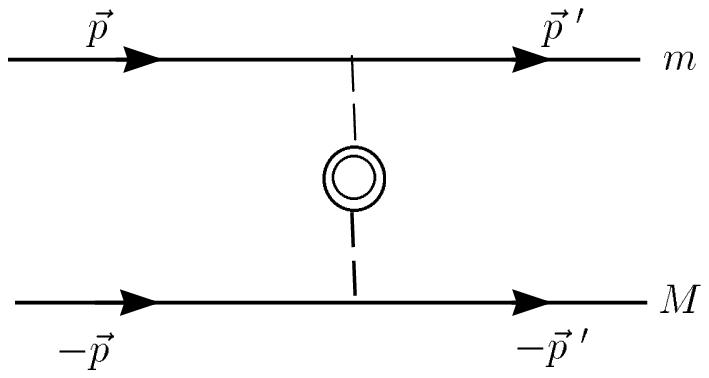}}
\caption{ }
\label{f-uehling}.
\end{figure}

To order $ \alpha^5$, and still in first order of perturbation theory, 
 we must either increase  $\kappa $ or $\sum n_i$  by one. It is
not possible to increase the number of inverse electron masses $\kappa$
 by one,
but there are two ways to increase $\sum n_i$ by one. 
One is to include
the one-loop corrections to the  coefficients $\delta$'s
of the vertices in Fig.[\ref{a4}],  so we
will have to match these interactions to one loop.
The other possibility for $\kappa =2$ and $\sum n_i =2$ is  to
consider the  new
 interaction  corresponding  to the vacuum polarization correction to
the Coulomb propagator which  is depicted in Fig.[\ref{f-uehling}].

\begin{figure}
\centerline{\epsfxsize 2.8 truein
\epsfbox {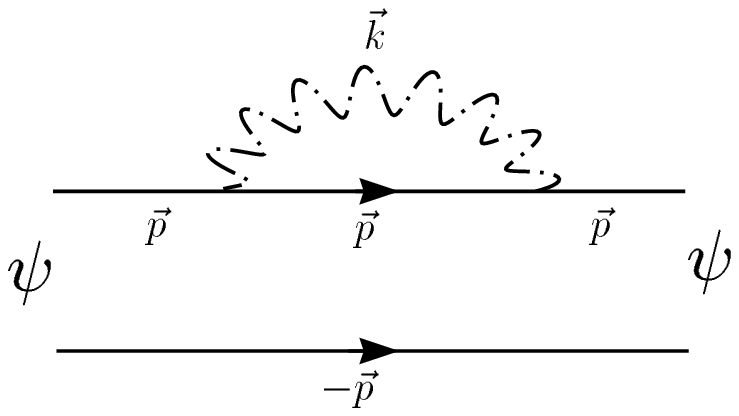}}
\caption{ }
\label{fultrasoft}.
\end{figure}

We now turn to diagrams in second order of perturbation theory, in which
case $N_{TOP} =1$. It can easily be verified that $\zeta$,
 Eq.(\ref{zeta}),
cannot then  be  made
equal to  $5$. We have now uncovered all the diagrams containing only
soft photons which contribute to order $\alpha^5$.
 The only remaining possibility is to consider diagrams with ultra-soft 
photons  which  lead to a contribution of the form given by 
 Eq.(\ref{crules}) but 
now with   (see Ref.[2]):
\beqa
\zeta=\sum_i n_i +1+\rho+\kappa-N_{top}+2N_\gamma+\sum_i {\cal M}_i,
\label{alphapower}
\eeqa
where $N_\gamma$ is the number of ultra-soft photons in the diagram and 
${\cal M}_i$ is the order, in the multipole expansion, of the $i_{th}$
vertex and the sum is over the vertices connected to ultra-soft photons
only.
 As before, we set $\zeta
=5$ and $\rho =0$ to obtain  non-recoil corrections  of order
$\alpha^5$. For diagrams containing ultra-soft photons, the minimum
value of $N_{TOP}$ is $1$, because these photons propagate in the
time direction (we again refer the reader to Ref.[2] for more
details).
Working
at the zeroth  order of the multipole expansion (${\cal M}_i=0$) and
considering  only one ultra-soft photon ($N_\gamma=1$),  
 we then  have the condition $\kappa + \sum n_i =3$. Since the ultra-soft
 photon is necessarily transverse and  transverse vertices contain at
least one power of inverse mass (see Fig.[1]),
$\kappa$ is
bigger or equal to $2$. In addition,  $\sum n_i$ is at least equal to one
({\it i.e.} there at least a total number of  one factor of $\alpha$ in
 the vertices). We therefore
 already fulfill the condition $\kappa + \sum n_i =3$
with the simplest diagram 
 which corresponds to an ultra-soft
photon connected to two $\vecp \cdot \vecA$ vertices, as represented in
Fig.[\ref{fultrasoft}].

We have now identified all diagrams contributing to the order of
interest. We now turn to the matching.
 From the above discussion, we see that we need to match
to one loop the coefficients of the interactions contributing
to order $\alpha^4$.

\begin{figure}
\centerline{\epsfxsize 5.2 truein
\epsfbox {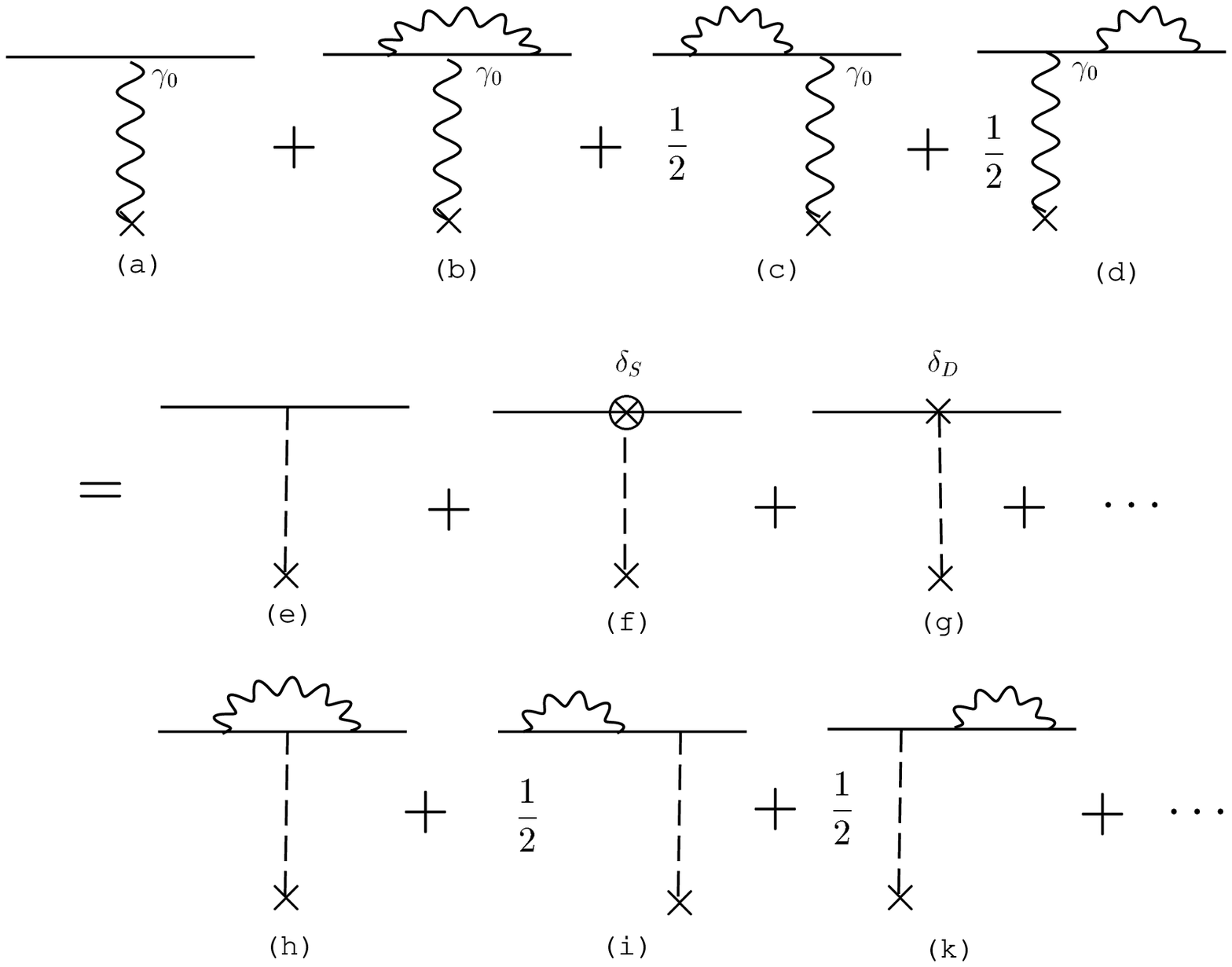}}
\caption{ }
\label{fmatching}.
\end{figure}

 To make NRQED agree with QED at the one loop order, we impose 
the relation  illustrated in Fig.[\ref{fmatching}]. This matching was
performed in \cite{kinoshita} but, even though our final result is
of course the same,  our derivation differs sufficiently to be 
 presented  it here.
  The QED scattering amplitude (the left hand side of
Fig.[\ref{fmatching}])  can be  expressed in terms of
 the usual form factors
$F_1 (Q^2)$ and $F_2 (Q^2)$ in the following way (to be consistent,
we use a nonrelativistic normalization for the Dirac spinors):
\beqa
&&-e {\bar u(\vecp~', \vecs') \over {\sqrt{2 E'}} }
\biggr[\gamma_0 \,  A^0(\vecQ) F_1(Q^2) 
-{i\over 2m}\sigma^{0j} \,  
 A^0(\vecQ)~Q^j~ F_2(Q^2) \biggl] \, { u(\vecp ,\vecs) \over {\sqrt{ 2E}}}
~~~{\Huge =} 
\nonumber\\
&&  F_1(Q^2)\xi^{' \dagger}
\biggr[-eA^0+{e\over 8m^2}\vecQ~^2 A^0 -{ie\over 4 m^2}\vecsigma 
 \cdot (\vecp~'\times\vecp)A^0 +\dots \biggl]
\xi +\nonumber\\
&&F_2(Q^2)\xi^{' \dagger}
\biggr[{e\over 4m^2}\vecQ~^2 A^0-{ie\over 2 m^2}\vecsigma 
\cdot (\vecp~'\times\vecp)A^0+\dots \biggl]
\, \xi, \label{left0}
\eeqa
 where $ \vecQ = \vecp' - \vecp $, $\vecs$ and $\vecs'$ are the initial and
final spin of the electron, respectively, and $\xi, \xi'$  are the
corresponding initial and final two component Pauli spinors normalized
to unity. We choose the convention that $e$ is positive so that the charge 
of the electron is $-e$. In (\ref{left0}), as in the rest of the paper, we use $m$ to
represent the mass of the electron.
 Notice that  the matching involves a double expansion. One
expansion is in 
the coupling constant 
$\alpha$ and  the  other is the nonrelativistic  expansion in $\vecQ/m$
which leads to renormalization of different NRQED operators.  
  The  nonrelativistic 
expansions of the form factors are
  \cite{kinoshita}
\beqa
F_1(Q^2)&= &1-{\a \over 3 \pi}\biggr[{\vecQ~^2 \over m^2}
\biggr(\ln({m \over 
\lambda})
-{3\over 8}\biggl) \biggl]+\dots \nonumber\\
 F_2(Q^2)&=& a_e - {\a \over \pi}{\vecQ~^2 \over 12 m^2}+\dots. 
\label{ff}
\eeqa
where $a_e$ is the electron anomalous magnetic moment which,
to the order of interest, can be taken to be $ \alpha/ (2 \pi)$. In the 
above Eqs., since $Q^0$ is of order of $v^2$ and $\vert\vecQ \vert$ is 
of order $v$, we have ignored $Q^0$ respect to $\vecQ$. 

 If we substitute (\ref{ff}) in (\ref{left0}) we obtain
 \beqa 
\xi^{' \dagger}  \,(-eA^0)\,  \xi~&+&{e\over 8m^2} \xi^{' \dagger}\,
\vecQ^2 A^0 \xi \biggl[ 1 + {8 \alpha \over 3 \pi}
\biggl( \ln ({m \over \lambda} )- {3 \over 8} \biggr) 
+ 2 a_e + \dots \biggr] \nonumber \\ && \quad \quad 
-{ie\over 4 m^2}
\xi^{' \dagger} \,\vecsigma \cdot (\vecp'\times\vecp)A^0  \,  \xi
\biggl[ 1 + 2 a_e + \dots \biggr] + {\cal O}( \vecQ^4/m^4).
\label{left}
\eeqa

We must now compute the right-hand side of Fig.[\ref{fmatching}] to
  complete the calculation of the one-loop renormalized NRQED 
coefficients.
Since we are dealing with ultra-soft photons in Figs. [\ref{fmatching}(h)],
[\ref{fmatching}(i)] and [\ref{fmatching}(k)],
 we  use the special Feynman rules  derived in  Ref.[2]. 
Working  at the zeroth order of the multipole expansion, 
  Fig.[\ref{fmatching}(h)]  corresponds to 
 \beqa
 \xi^{' \dagger} ({2e  p_i \over 2m})
({2e  p'_j \over 2m})\!\!\!\!\!\!\! &&\int{ d^3k\over (2\pi)^3}
 {\d_{ij}-{k_ik_j\over \veck^2+\lambda^2}
\over 2 \sqrt{\veck^2+\lambda^2}}~{1\over   -\sqrt{\veck^2+\lambda^2}}~(-eA_0)~
{1\over{\vecp^2\over2m}-{\vecp'^2\over 2m}-\sqrt{\veck^2+\lambda^2}}
~\xi\nonumber\\
&\approx& \xi^{' \dagger }(-eA_0) \xi \,
({e\over m})^2 ~{\vecp' \cdot \vecp\over 3} 
\int{ dk~ \veck^2\over 2\pi^2}~~{2\veck^2+3\lambda^3\over \veck^2+\lambda^2 }
~{1\over 2 \sqrt{\veck^2+\lambda^2}}{1\over \veck^2+\lambda^2}\nonumber\\
&=& - {8 \alpha \over 3 \pi}
 \biggl[\ln ({2\Lambda \over \lambda} )- {5 \over 6} \biggr] 
 \, \xi^{' \dagger} eA_0 \xi \,
{ \vecp \cdot \vecp' \over 4 m^2}  \label{nineh}
 \eeqa
where, in the second line, we have used the fact that the integral
is already proportional to $\vecp \cdot \vecp'$ to approximate
$\vecp^2 \approx \vecp'^2 \approx 0$ in the propagators. The
corrections to these expressions will lead to higher order
operators.

Notice  that we are only  working with scattering diagrams  when
performing the 
 matching, no bound state physics enters this stage of the calculation.
 To compute the amplitudes in Figs.[\ref{fmatching}(i)] and
 [\ref{fmatching}(k)], we just need to evaluate the first
one and  then obtain the second one by replacing  $\vecp$ by $\vecp~'.$ 
For  Fig.[\ref{fmatching}(i)], we can write
\beqa
{1\over2} \xi^{' \dagger} ({2e p_i \over 2m})
({2e  p_j \over 2m})\int { d^3k\over (2\pi)^3}
 {\d_{ij}-{k_ik_j\over \veck^2+\lambda^2}
\over 2 \sqrt{\veck^2+\lambda^2}}~{1\over E-{\vecp^2\over 2m}
 -\sqrt{\veck^2+\lambda^2}}~{1\over E-{\vecp^2\over 2m}}~(-eA_0)~\xi.
\nonumber\\ \label{patr}
\eeqa 
Here, $E$ represents the on-shell energy $\vecp^2 /2 m$.
Of course, in that limit, the propagator $1/(E- \vecp^2 / 2m)$
is divergent, which signals the need for a mass renormalization.
In NRQED, we perform mass renormalization exactly  as in QED, {\it
i.e.} we start by keeping $E \neq \vecp^2/ 2m$  and subtract
the mass counter-term:
\beqa
-{1\over2} \xi^{' \dagger}eA_0({e\over m})^2 ~{\vecp^2\over 3} 
 \!\!\!\!\!\!&&\int{ dk~\veck^2\over 2\pi^2}~~{2\veck^2+3\lambda^3\over \veck^2+\lambda^2 }
~{1\over 2 \sqrt{\veck^2+\lambda^2}} 
\nonumber\\
&&\biggl( {1\over E-{\vecp^2\over 2m}- \sqrt{\veck^2+\lambda^2}}-
~{1\over -  \sqrt{\veck^2+\lambda^2}} \biggr)~{1\over E-{\vecp^2\over 2m}}
~\xi.
\eeqa
Expanding the term in the parenthesis around  $E-{\vecp^2\over 2m}$,
we get a series which, by construction, starts at order $(E- \vecp^2/
2m)^1$, which cancels the propagator $1/(E- \vecp^2/2m)$ in
Eq.(\ref{patr}). One can then finally take the limit $E \rightarrow
 \vecp^2/2m$ with for result, for the sum of 
 Figs.[\ref{fmatching}(i)]
 and [\ref{fmatching}(k)],
\beqa
-{1\over 2} \xi^{' \dagger}~eA_0 ~{\a \over 3 \pi m^2}(\vecp^2+\vecp'^2)\int dk \veck^2
{2\veck^2+3\lambda^2\over \veck^2+\lambda^2}~{1\over \sqrt{\veck^2+\lambda^2}} 
~{-1\over \veck^2+\lambda^2}\xi\nonumber\\ 
= -{8 \alpha \over 3 \pi}
 \biggl[\ln ({2\Lambda \over \lambda} )- {5 \over 6} \biggr] \biggl({
 - \, \xi^{' \dagger} eA_0
\xi \over 8 m^2}(\vecp^2+\vecp'^2)\biggr). \label{ninei}
\eeqa
Putting everything together, the complete right-hand side of 
Fig.[\ref{fmatching}] is equal to the sum of Eqs.(\ref{nineh}) and
(\ref{ninei}): 
 \beqa
\xi^{' \dagger} \,(-eA^0)\, \xi~ &+&{e\over 8m^2} \xi^{' \dagger}\,
\vecQ^2 A^0 \xi \biggl[  \d_D+{8 \alpha \over 3 \pi}
\biggl( \ln ({2\Lambda \over \lambda} )- {5 \over 6} \biggr) 
  \biggr] \nonumber \\
&-& \d_S \,  {ie\over 4 m^2}
\xi^{' \dagger} \,\vecsigma \cdot (\vecp~'\times\vecp)A^0  \,  \xi
 \label{right}
\eeqa 

Now,  by equating  (\ref{left}) and (\ref{right}), we can
evaluate  $\d_D$ and $\d_S$:
\beqa
\delta_D ~&=&~ 1 + {\alpha \over \pi} {8 \over 3} \biggl[ \ln 
\bigr( { m
\over 2 \Lambda} \bigr) + {11 \over 24} \biggr] + 2 a_e  
+ {\cal O} ( \alpha^2)     \label{deltaS}\\ 
\delta_{S} ~&=&~ 1 + 2 a_e 
+ {\cal O} ( \alpha^2) \label{deltaD} . 
\eeqa

Notice that, since the  relativistic kinetic vertex does not enter
the matching, we also find\footnote{One might expect that $\delta_R$
be equal to one to all orders, since this interaction comes from Taylor
expanding the relativistic expression for the energy, but this might not
be so if the regulator (as is the case in this calculation)
 breaks Lorentz invariance.}
\beqa
\delta_R=  1 +{\cal O} (\alpha^2) .
\eeqa
We are now ready for the third and last step of  the calculation,
the computation of the bound state diagrams {\it per se}.
Since only  $\delta_D$  and $\delta_S$ receive an ${\cal O} (\alpha)$
correction, but not $\delta_R$,   
 only  Figs.[\ref{a4}(b)] and [\ref{a4}(c)] are needed for the ${\cal O}
(\alpha^5)$ calculation. 
Let us now start with Fig.[\ref {a4}(b)]. 
\beqa 
\Delta E_{7(b)} &= & \d_S{ -iZ e^2 \over 4 m^2} \int { d^ 3p' d^3p 
\over (2 \pi)^6} \Psi^*(\vecp)
\biggl( { \vecsigma \cdot \vecp~' \times \vecp \over 
(\vecp~' -\vecp)^2}
\biggr) \Psi(\vecp~')\nonumber\\
&= &\d_S~ { Z\a \over 2 m^2}<{ \vecs \cdot \vecL\over r^3} >
\nonumber \\   
&= &\d_S~{m( Z \a)^4 \over 4  n^3 (l+1/2)}
\biggr({\d(j-(l+1/2))\over (l+1)}-
{\d(j-(l-1/2))\over l}\biggl)(1-\d_{l,0}),\nonumber\\   
 \label{spinorbit}
\eeqa
where $\Psi(\vecp)$ is  the
 Schr\"odinger wavefunction (including the electron spin\footnote{
In the non-recoil limit, the spin of the proton
completely decouples from the problem.} corresponding to the quantum
numbers $n,j$ and $l$.
 
In Eq.(\ref{spinorbit})  we  used 
\beqa
\int{ d^3p'\over (2\pi)^3}e^{i\vecp~'\cdot \vecr~'}
{\vecsigma\cdot\vecp \times \vecp~'\over \vecp'^2}= 
i {\vecsigma\cdot\vecp \times \vecr~'\over4 \pi r'^3} .
\eeqa

For the diagram of Fig.[\ref{a4}(c)], we  find
\beqa
\Delta E_{7(c)}
 &= & \int { d^ 3p' d^3p \over (2 \pi)^6} \Psi^*(\vecp~')
 \biggl(
\d_D~ { e 
(\vecp~'
 - \vecp)^2 \over 8 m_a^2}~Ze~{ 1 \over (\vecp~' - \vecp)^2} 
 \biggr)
\Psi(\vecp)
\nonumber\\
  &= & \int { d^ 3p' d^3p \over (2 \pi)^6} \Psi^*(\vecp~')
\Psi(\vecp)\nonumber\\
&=&\d_D~{Ze^2\over8 m^2}\vert
 \Psi(0)\vert^2=\d_D~{m(Z \a)^4 \over2 n^3}\d_{l,0}.\label{darwin} 
\eeqa

Using the counting rules, we also found that the diagram depicted
in  Fig.[\ref{f-uehling}] would contribute to ${\cal O}( m \alpha^5)$.
 This diagram
corresponds to the well-known Uehling potential  and is found to be
\beqa
\D E_8 &= & \int{ d^3p'd^3p \over (2\pi)^3}\Psi^*(\vecp~')
\biggr( -e~{1\over 
(\vecp~'-\vecp)^2}
 \d_{VP}{-(\vecp~'-\vecp)^4 \a \over15 \pi m^2}~{1\over 
(\vecp~'-\vecp)^
2}~Ze~\biggl)\Psi (\vecp) \nonumber \\
&=& -\d_{VP}Ze^2{\a \over 15 \pi}{1\over m^2}\vert \Psi(0)\vert^2  
= - \d_{VP}{4 \a \over 15 \pi}{m(Z \a)^4\over n^3}\d_{l,0}.  \label{uehling}
\eeqa

We finally turn our attention to the only remaining diagram, which
is represented in Fig.[10]. 
 The corresponding integral is (as shown in Ref.[2], 
 in  zeroth order 
of the multipole expansion we set $\vecp~'= \vecp$ on the vertices):
\beqa
\Delta E_8 &= & \int { d^ 3k d^3p \over (2 \pi)^6} \Psi^*(\vecp) ~
 { 2 e   p_i   \over  2 m}   {2 e   p_j   \over 2  m}  
  { 1 \over 2k }{1\over E_n-{\vecp^2\over 2m}-k}
 ( \delta_{ij} - {   k_i k_j\over \veck^2}) ~ \Psi (\vecp)\nonumber \\
&= & {e^2\over 2m^2(2 \pi)^3}\int {  d^3p \over (2 \pi)^3}
\Psi^*(\vecp)\int dk~k d\Omega      
 { (\vecp^2 - {  (\vecp \cdot \veck)^2\over \veck^2})\over E_n-
{\vecp^2\over 2m}-k}
   \Psi (\vecp)\nonumber \\
&= & { 2\a \over3 \pi} 
 \int{ d^3p\over (2 \pi)^3}\int dk \, k ~ \Psi^*(\vecp) ~  
 ({\vecp^2/m^2\over E_n-{\vecp^2\over 2 m}-k})~\Psi (\vecp). 
\label{uu}
\eeqa
In a bound state, beyond tree level  
   one must include an
 infinite number of Coulomb lines  in the intermediate state.
This can easily be seen from the counting rules,  Eq.(\ref{alphapower}).
 Indeed,  adding a
 Coulomb line will not change
  $\zeta$  because this  increases both
 $N_{top}$ and
$\sum_i n_i$ by one, which has no overall effect. Because of this,
one must use the Coulomb Green's function for the
intermediate state. Using the bra and ket notation, Eq.(\ref{uu})
 must then be replaced by the well-known expression:
\beqa
{2 \a \over 3 \pi}\sum_{n'} \int dk~k~{<n| \vecv_{op}|n'>
<n'| \vecv_{op}|n>
 \over E_n-E_{n'}-k}. 
\eeqa 
This part of the calculation is well known and is carried out 
in many texbooks (see for example Ref.\cite{itzykson}). The result
is
\beqa
 \D E=  m  {4 \a\over 3 \pi} {(Z\a)^4\over n^3}
\left\{ \begin{array}{ll} ~~\ln {\Lambda \over<E_n>}~~~ & 
\mbox{if $ l= 0$}\\ \\
  ~~\ln {Z^2 m\a^2 \over 2<E_n>}~ ~~ & \mbox
{if $l \neq0$},
\end{array} \right.  \label{ultrasoft}
 \eeqa
where $<E_n>$ is the Bethe logarithm which can be evaluated numerically
 \cite{yennie}. Now, using Eqs.(\ref{deltaS})  and (\ref{deltaD}), 
 we add the $\a^5$ contributions from 
Eqs.(\ref{spinorbit}) and(\ref{darwin})
to Eqs.(\ref{uehling}) and (\ref{ultrasoft}) to   obtain
\beqa
\D E=m {4\a \over 3 \pi}{(Z\a)^4 \over n^3} \left\{ \begin{array}{ll}
 ln {m  \over2<E_{n,0}>}+{19\over 30}~~~ & \mbox{if $ l= 0$}\\ \\

 ~~\ln {Z^2 m \a^2 \over 2<E_{n,l}>}+{3\over8(2l+1)}({\d(j-(l+1/2))\over
 (l+1)}-
{\d(j-(l-1/2))\over l}) ~ ~~ & \mbox
{if $l\neq0$},
\end{array} \right.
\eeqa
which is  the well-known Lamb shift.

 \section{Conclusion}
We have calculated the complete ${\cal O} ( m \alpha^5)$ 
non-recoil corrections to the hydrogen energy levels, also
referred to as the Lamb shift. The superiority of    NRQED  over the
traditional approaches
 is twofold. Firstly, the calculation of the bound
state diagrams is greatly simplified because the use of an effective
field theory permits to disentangle the contributions from low
and high momenta and only    QED {\it scattering} diagrams
need to be evaluated. Secondly, the NRQED calculation is {\it
systematic} in the sense that simple counting rules can
be used to isolate the diagrams contributing to a given
order in $\alpha$, which is not possible in traditional approaches.
\begin{figure}
\centerline{\epsfxsize 5.2 truein
\epsfbox {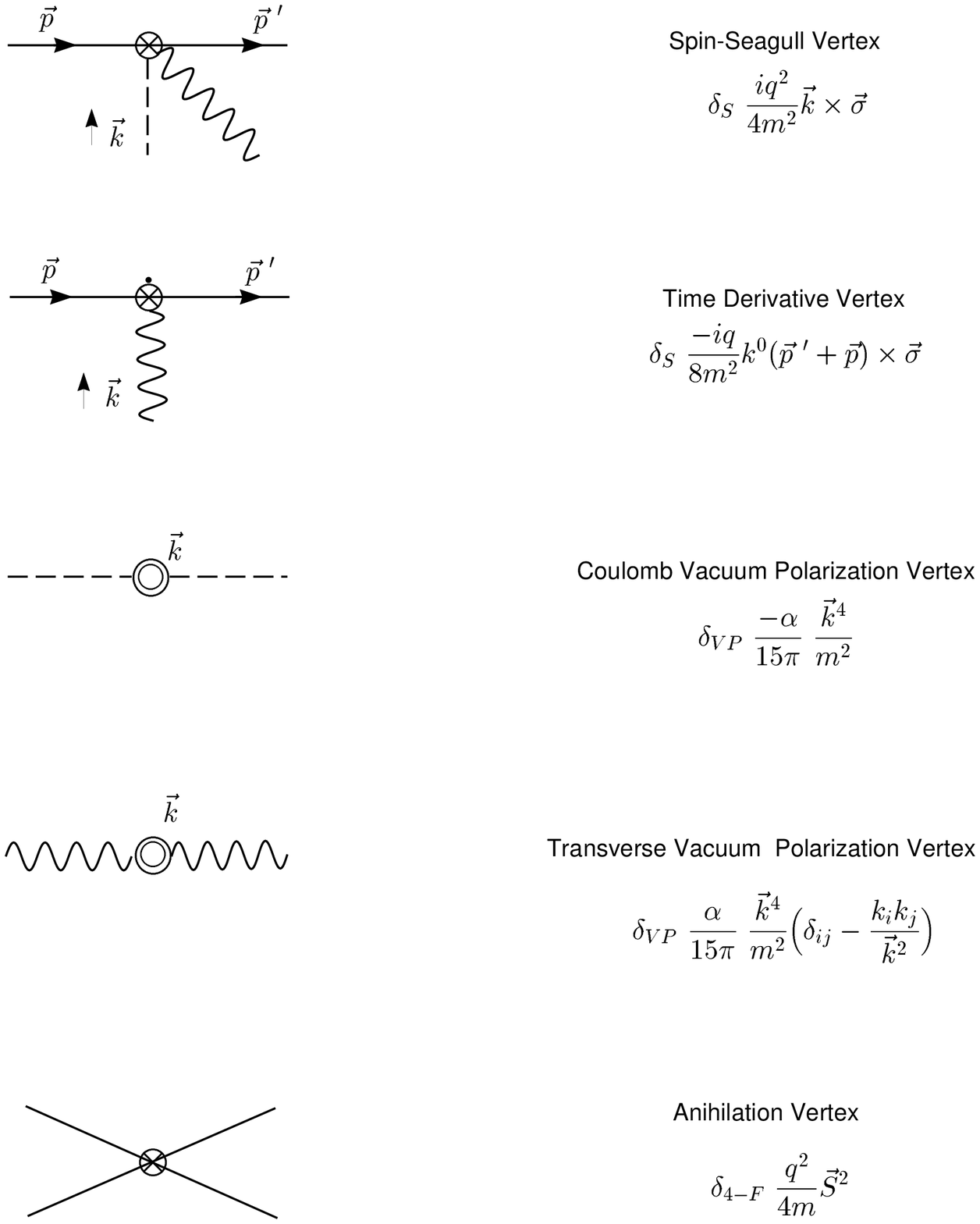}}
\centerline{NRQED Vertices (continued)}
\end{figure}


\end{document}